\documentclass{article}
\usepackage{amssymb}

\usepackage{graphicx}
\usepackage{amsmath}

\newcommand{\nwc}{\newcommand}
\nwc{\bc}{{\bf{C}}}
\nwc{\br}{{\bf{R}}}
\nwc{\bz}{{\bf{C}}}

\begin{document}

\
\bigskip

\begin{center}
{\large \textbf{GEOMETRY OF MIXED STATES AND DEGENERACY STRUCTURE OF
GEOMETRIC PHASES FOR MULTI-LEVEL QUANTUM SYSTEMS. A UNITARY GROUP APPROACH} }

\vskip1cm

\noindent E. Ercolessi$^{1}$, G. Marmo$^{2}$, G. Morandi$^{1}$, N. Mukunda$%
^{3,4}$ \vskip0.5cm \noindent \textit{{$^{1}$ Dipartimento di Fisica,
Universita' di Bologna, INFN and INFM,\\[0pt]
Viale Berti-Pichat 6/2, 40127 Bologna, Italy\\[0pt]
$^{2}$ Dipartimento di Scienze Fisiche, Universita' di Napoli Federico II
and INFN, \\[0pt]
Via Cinzia, 80126 Napoli, Italy\\[0pt]
$^{3}$ Centre for Theoretical Studies and Department of Physics, \\[0pt]
Indian Institute of Science, Bangalore 560 012, India\\[0pt]
$^{4}$ Jawaharlal Nehru Centre for Advanced Scientific Research,\\[0pt]
Bangalore 560 064, India } }
\end{center}

\vskip2cm Short title:\textit{Geometric Phases for Multi-Level Quantum Systems}\vskip1cm PACS: 02.20;
03.65\vskip2cm
\begin{abstract}
We analyze the geometric aspects of unitary evolution of general states for a
multilevel quantum system by exploiting the structure of coadjoint orbits
in the unitary group Lie algebra. Using the same methods in the case of $%
SU(3)$ we study the effect of degeneracies on geometric phases for three-level systems. This is shown to lead to
a highly nontrivial generalization of the result for two-level systems in which degeneracy results in a
"monopole" structure in parameter space. The rich structures that arise are related to the geometry of adjoint
orbits in $SU(3)$. The limiting case of a two-level degeneracy in a three-level system is shown to lead to the
known monopole structure.
\end{abstract}
\newpage

\section{Introduction.}

\bigskip

The original discovery and development of the geometric phase ideas were in
the context of cyclic evolution within the Adiabatic Theorem of Quantum
Mechanics \cite{Be}. In this treatment the adiabatic time dependence of the
Hamiltonian operator was supposed to arise from its dependence on classical
external parameters which in turn were taken to be slowly varying functions
of time. Thus the quantum geometric phase was associated with a closed
circuit in the external parameter space rather than in the Hilbert or ray
space of the quantum system.

Subsequent work has shown that the geometric phase can be viewed as belonging intrinsically to the ray space of
the quantum system, and no reference need be made to the space of classical external parameters \cite {SW,AA}.
The geometric phase idea has been applied also in the study of dissipative quantum-mechanical and classical
dynamical systems \cite{La}. However the original picture shows its usefulness in certain physical situations
and should therefore be retained. In particular the parameter space is most appropriate for studying the effects
of degeneracies in the eigenvalue spectrum of the quantum Hamiltonian. In his original work \cite {Be}, Berry
studied the most important practical case of two-level degeneracy. Near such a point in parameter space, with no
essential loss of generality one can suppose that there are generically three independent real parameters in the
Hamiltonian. In accordance with the von Neumann-Wigner theorem it requires the vanishing of all three parameters
to produce a degeneracy of levels in the Hamiltonian. In this three-dimensional picture the point of degeneracy
appears as a ``magnetic monopole" (in parameter, not in physical space), and this singular point structure
dominates the geometric phases for closed circuits in its vicinity.

The aim of this paper is to discuss some general geometric aspects of the
unitary evolution of a quantum system with a finite number of levels,
specializing later to a detailed study of the generic features of a
three-level system, which would be the next level of complication in the
quantum mechanical sense after two-level systems. For this problem,when
degeneracies occur, the earlier ``magnetic monopole'' in the geometric phase
discussion gets replaced by a far richer singularity structure in an
eight-dimensional parameter space, and essentially new geometric as well as
algebraic features arise. The situation hardly looks like a ``monopole'' any
more. The basic reason is that while the treatment of two-level degeneracy
involves the group $SU(2)$ and its (real three-dimensional) adjoint
representation, with three active levels one has to work with the group $%
SU(3)$ and its (real eight-dimensional) adjoint representation. We make contact here with some previous work on
$SU(3)$ \cite{AC,Ar,MM}, putting however more emphasis on the geometric aspects involved in the analysis.
The more general framework in which we study the occurrence of geometric phases and that is presented here can
be of some relevance to studies and experiments on quantum and/or optical systems with a finite number of
degrees of freedom that can be described on finite-dimensional Hilbert spaces, such as those occurring in
many-channel (and three-channel in particular) optical interferometry \cite{RZ,San,San2}, quantum computing
\cite{S,Sc,Za,Ch} or the study of entangled states \cite{Ent,Sc,HW,Ni,DC,BP} of simple composite systems.

The material of the paper is organized as follows. In Sect.$2$ we discuss
some general features of a generic $n$-level quantum system as described, in
the von Neumann picture, in the space of mixed states (or density matrices)
and analyze the structure of the orbits under the action of the appropriate
unitary group, i.e. $U\left( n\right) $. In Sect.$3$ we collect the basic
features of a Hermitian $3$-level Hamiltonian, its eigenvalues and
eigenvectors and the regions of double and triple degeneracy in parameter
space. We make use of the defining and adjoint representations of $SU(3)$
and whenever possible bring in the $SU(3)$ invariants of the problem. In
Sect.$4$ we disscuss briefly the geometric properties of adjoint orbits in
the Lie algebra of $SU(3)$. Sect.$5$ deals with the three independent
geometric phase two-forms, relates them to the symplectic structures on the
orbits, and develops a ''sum rule'' for them. The next Sect.$6$ analyzes
these two-forms from the point of view of their $SU(3)$ transformation
behaviour and shows that there are three independent tensor contributions
belonging respectively to the octet $($ $\underline{8})$, decouplet $(%
\underline{10})$ and $\ $antidecouplet $(\underline{10}^{\star })$
irreducible representations of $SU(3)$. The details of recombining these
three parts and reconstituting the complete two-forms are given. Sect.$7$
studies the limiting case of a two-level degeneracy in a three-level system
and traces in detail the emergence of the more familiar ``monopole''
structure as a special case in the present formalism. Sect.$8$ contains some
concluding remarks, and in the Appendices some relevant materials related to
$SU(3)$ and its irreducible tensors, as needed here, are collected.

\bigskip

\section{General Aspects for $n$-level Systems.}

\bigskip

Many aspects of quantum dynamical systems are described on the space of
mixed states (also called density matrices) where the evolution is ruled by
the von Neumann equation:
\begin{equation}
i\hbar \frac{d\rho }{dt}=[H,\rho ]  \label{vn}
\end{equation}
where $\rho $ describes the mixed state and $H$ is the (time-independent)
Hamiltonian of the system. The von Neumann equation may be obtained from the
Schr\"{o}dinger equation:
\begin{equation}
i\hbar \frac{d}{dt}|\psi \rangle =H|\psi \rangle  \label{se}
\end{equation}
with $\rho $ being defined as
\begin{equation}
\rho =\sum\limits_{j}c_{j}|\psi _{j}\rangle \langle \psi _{j}|,\text{ }%
c_{j}\in \br^{+}  \label{rho}
\end{equation}
and the condition: $Tr\rho <+\infty $.

The dynamical evolution given by integrating the von Neumann equation has
the form:
\begin{equation}
\rho(t)=\exp[-iHt/\hbar]\rho(0)\exp[iHt/\hbar]  \label{rhot}
\end{equation}
i.e. it is a conjugation in the space of mixed states.

In many physical situations (e.g. in many-channel interferometry and \
quantum computing \cite{Sc,Za,Ch} we deal actually with finite-level quantum
systems, in which case the Hilbert space of states is given by $\mathcal{H}=%
\bc^{n}$ for some $n$, and the group of unitary transformations is the
unitary group $U(n)$. The orbits of the group, acting on the space of mixed
states, will determine the carrier spaces of the dynamical evolution
according to the von Neumann equation, i.e. each initial condition will
select an orbit to which it belongs and the evolved state will remain on the
same orbit. If instead a dissipative term is added on the r.h.s. of the von
Neumann equation, in general the evolution will not be unitary anymore and
the initial state will be carried from one orbit to another depending on the
specific dissipation mechanism. These remarks point to the fact that the
total space of the orbits may become relevant when dissipation occurs.

Here we are not going however to deal with particular dynamical systems but
would rather like to concentrate our attention on the ``carrier spaces'',
i.e. we are emphasizing the kinematical rather than the dynamical aspects.
In particular we would like to consider the space of mixed states (convex
combinations of pure states) and study how they are ``partitioned'' into
orbits of $U(n)$ under conjugation. Eventually, in paving the way to the
comparison of the von Neumann and the Schr\"{o}dinger equations we will
consider the connection one-form that is usually associated with the
``geometric'' or ``Berry'' phase.

The basic idea of our approach consists in embedding the space of mixed
states in the space of all Hermitean matrices, i.e. instead of considering
only convex combinations of pure states we will enlarge to combinations with
arbitrary real coefficients. Hereafter we will multiply them by the
imaginary unit $i$ so that they will define the Lie algebra $u(n) $ of the
unitary group. With this extension our analysis will reduce to the analysis
of the adjoint (or coadjoint) orbits of $U(n)$ on its Lie algebra.

The structures available on the space of matrices will allow us to consider
bilinear products that are ``inner'' in $u(n)$, and this will provide us
with computational tools making the analysis of the orbits more manageable.
In the next Sections more detailed computations will be carried out for a
three-level quantum system, i.e. for $U(3)$. In this particular context we
will also exhibit the connection one-form or ``Berry connection''.

\bigskip

\subsection{Algebraic Structures on the Enlarged Space of Mixed States.}

\bigskip

Given $\mathcal{H}=\bc^{n}$, we consider an orthonormal basis: $%
\{|\psi_{i}\rangle\}_{i=1}^{n},\langle \psi_{i}|\psi_{j}\rangle =\delta_{ij}$%
, and construct a basis $\{E_{ab},E_{ab}^{\prime}\},$ in the space of
Hermitean matrices as:
\begin{eqnarray}
E_{aa}&=&|\psi_{a} \rangle \langle\psi_{a}|  \label{b1} \\
E_{ab}&=&|\psi_{a} \rangle \langle\psi_{b}|+ |\psi_{b} \rangle
\langle\psi_{a}|, \; a < b=1,...,n
\end{eqnarray}
and
\begin{equation}
E_{ab}^{\prime}=i\{|\psi_{a} \rangle\langle\psi_{b}|-|\psi_{b}
\rangle\langle\psi_{a}|\},\; a< b=1,...,n  \label{b2}
\end{equation}
Any Hermitean matrix $H$ can be written in the form:
\begin{equation}
H=\xi^{ab}E_{ab}+\xi^{\prime ab}E_{ab}^{\prime}  \label{herm}
\end{equation}
with real coefficients.

In this real vector space we may consider two bilinear products, namely:
\begin{equation}
H_{1}\ast H_{2}=\frac{1}{2}\{H_{1}H_{2}+H_{2}H_{1}\}  \label{star}
\end{equation}
which is commutative but not associative, and:
\begin{equation}
H_{1}\wedge H_{2}=i\{H_{1}H_{2}-H_{2}H_{1}\}  \label{wedge}
\end{equation}
which defines a Lie algebra structure. Also, a scalar product is inherited
from that in $\mathcal{H}$, namely:
\begin{equation}
H_{1}\cdot H_{2}=Tr\{H_{1}H_{2}\}  \label{scalar|}
\end{equation}

The specific way we have written our bilinear structures permits us to write
down several identities coming from the ``interaction'' of these products.
We list below some of them:

\begin{itemize}
\item[i)]  The scalar product is invariant under conjugation:
\begin{equation}
H_{1}\cdot H_{2}=U(H_{1})\cdot U(H_{2})  \label{inv}
\end{equation}
where
\begin{equation}
U(H)=U^{\dagger }HU,\text{ \ }U\in U(n)  \label{conju}
\end{equation}

\item[ii)]  If $H$ is generic, the set of powers $%
H^{0},H^{1},H^{2},...,H^{n-1}$ defines a maximal set of commuting Hermitean
matrices. This generates the \textbf{commutant} of $H$, and it is Abelian.
When $H$ is not generic, i.e. there are degeneracies in its spectrum, not
all powers are independent, and the independent ones are in a number equal
to the degree of the minimal polynomial associated with $H$. The commutant
of $H$ will be no more Abelian and will have dimension greater than $n$.

\item[iii)]  Two Hermitean matrices will be in the same orbit iff they have
the same characteristic polynomial. We recall that the characteristic
polynomial $P(\lambda )$ is defined as:
\begin{equation}
P(\lambda )=\lambda ^{n}+c_{1}\lambda
^{n-1}+...+c_{n}=(-1)^{n}Det\{H-\lambda \mathbf{I}\}  \label{pol}
\end{equation}
where:
\begin{equation}
c_{1}=-TrH;\text{ \ }c_{k}=-\{Tr(H^{k})+c_{1}Tr(H^{k-1})+...+c_{k-1}TrH\},%
\;k\geq 2  \label{coeff}
\end{equation}

\item[iv)]  Two orbits will be \textbf{of the same type} iff their
commutants (seen as subalgebras of $u(n)$) are isomorphic.

\item[v)]  Each orbit is a symplectic manifold and is symplectomorphic with $%
U(n)/U_H$, where $H$ is an element of the orbit and $U_H$ its stabilizer
under conjugation, i.e.:
\begin{equation}
U_H=\{U\in U(n)\vdash U^{\dagger }HU=H\}  \label{stab}
\end{equation}
For a generic $H$:
\begin{equation}
U_H=\underset{n\text{ }times}{\underbrace{U(1)\times U(1)\times ...\times
U(1)}}  \label{form}
\end{equation}
At the other extreme, if $H$ corresponds to a pure state projection, $%
H=|\psi \rangle \langle \psi |$, then: $U_H=U(1)\times U(n-1)$. The space
of pure states is therefore given by:
\begin{equation}
\mathcal{P(H)}=\frac{U(n)}{U(1)\times U(n-1)}\approx \frac{SU(n)}{U(1)\times
SU(n-1)}  \label{pure}
\end{equation}

\item[vi)]  On each orbit there is a symplectic structure\cite{bal1,bal2} given by:
\begin{equation}
\omega _{H}=Tr\{HdU^{\dagger }\wedge dU\}  \label{sympl}
\end{equation}
where $U\in U(n).$ The way $\omega _{H}$ has been written defines actually a
two-form on the whole of $U(n)$, which is however degenerate. It is not
difficult to show that the kernel of $\omega _{H}$ coincides with the Lie
algebra of $U_H$. Therefore $\omega _{H}$ ``descends to the quotient'' and
defines a nondegenerate two-form on the orbit.

\item[vii)]  On $U(n)$ $\omega _{H}$ is exact, and is given by\cite{bal1,bal2}:
\begin{equation}
\omega _{H}=dTr\{HU^{\dagger }dU\}=-Tr\{HU^{\dagger }dU\wedge U^{\dagger
}dU\}  \label{usympl}
\end{equation}
The one-form $Tr\{HU^{\dagger }dU\}$ however does not descend to the
quotient, implying that on each orbit the symplectic form is closed and
nondegenerate, but not exact.

We recall that if $G$ is any Lie group and $X_{A},X_{B}$ are the
left-invariant vector fields associated with $A,B$ in the Lie algebra $g$ of
$G$, $\imath _{X_{A}}(U^{\dagger }dU)=A$ and $\imath _{X_{B}}(U^{\dagger
}dU)=B$, then:
\begin{eqnarray}
&&\omega (H)(X_{A},X_{B})=-Tr\{H[A,B]\}  \label{vec} \\
&=&-Tr\{HU^{\dagger }dU(X_{A})U^{\dagger }dU(X_{B})\}+Tr\{HU^{\dagger
}dU(X_{B})U^{\dagger }dU(X_{A})\}  \notag
\end{eqnarray}
Then:
\begin{equation}
\omega (H)(X_{A},X_{B})=Tr\{H[A,B]\}=Tr\{[B,H]A\}=Tr\{[H,A]B\}  \label{vec1}
\end{equation}
and it is clear that the kernel of $\omega $ coincides with the vector fields generated by the commutant of $H$.
Going to the quotient we obtain the symplectic structures on the orbits. In the case of $SU(3)$ an explicit
expression in terms of the Euler angles is given in \cite{By,By2,BBSS}.
\end{itemize}

\bigskip

\section{Structure of the Hamiltonian, Spectrum and Degeneracies for
Three-Level Systems.}

\bigskip

For a quantum system in which three (generically) distinct energy
eigenvalues and eigenvectors play a dominant role and the other states may
be neglected, the generic Hamiltonian is some Hermitean three-dimensional
matrix. With no loss of generality and whenever needed, by subtracting an
appropriate multiple of the identity, we may remove the trace of the matrix,
so that at the point of triple degeneracy all three eigenvalues vanish. We
recall that the set of all Hermitian $3\times 3$ matrices defines the Lie
algebra $u(3)$ of the unitary group $U(3)$, while the traceless ones define
the Lie algebra $su(3)$ of $SU(3)$.

A basis of generators for $su(3)$ is given by the (Hermitean and traceless) Gell'Mann matrices \cite{Ge,deS} of
Appendix A.

A generic $3\times3$ Hermitean matrix can be uniquely represented in the
basis of the Gell'Mann generators \cite{Ge} as:
\begin{equation}
H(\xi)=\xi_{0}\lambda_{0}+\frac{1}{2}\vec{\xi} \cdot \vec{\lambda}
\label{basis}
\end{equation}
where: $\lambda_{0}=\mathbf{I}$ is the identity matrix, $\xi=(\xi _{0},\vec{%
\xi})$ , $\xi_{0}=\frac{1}{3}Tr\{H\}$ and $\vec{\xi}$ is an
eight-dimensional real vector, $\vec{\xi}\in \br^{8}$. Of course: $\xi_{0}=0$
for traceless matrices (and viceversa).

Under conjugation with $U\in SU(3)$, $H$ transforms as:
\begin{equation}
UH(\xi )U^{\dagger }=\xi _{0}\lambda _{0}+\frac{1}{2}\vec{\xi}\cdot U%
\overrightarrow{\lambda }U^{\dagger }=:\xi _{0}\lambda _{0}+\frac{1}{2}\vec{%
\xi ^{\prime }}\cdot \vec{\lambda}  \label{tr}
\end{equation}
where (see formula ({\ref{transf})):
\begin{equation}
\xi _{r}^{\prime }=D_{rs}(U)\xi _{s}  \label{octet}
\end{equation}
}

The bilinear products discussed in Sect.$2.1$ induce binary operations among
vectors in $\br^{8}$. If: $H_{1}=H(\xi^{1}),$ $H_{2}=H(\xi^{2})$, then

\begin{itemize}
\item[i)]
\begin{equation}
H_{1}\wedge H_{2}=(\vec{\xi ^{1}}\wedge \vec{\xi ^{2}})\cdot \vec{\lambda}
\label{bin1}
\end{equation}
where:
\begin{equation}
(\vec{\xi ^{1}}\wedge \vec{\xi ^{2}})_{r}=-\frac{1}{2}f_{rst}\xi _{s}^{1}\xi
_{t}^{2}  \label{with1}
\end{equation}

\item[ii)]
\begin{eqnarray}
H_{1}\ast H_{2} &=&\frac{1}{3}\lambda _{0}Tr\{H_{1}H_{2}\}+\frac{1}{2}\{\xi
_{0}^{1}\vec{\xi ^{2}}+\xi _{0}^{2}\vec{\xi ^{1}}\}\cdot \vec{\lambda}
\notag \\
&+&\frac{1}{4\sqrt{3}}(\vec{\xi ^{1}}\ast \vec{\xi ^{2}})\cdot \vec{\lambda}
\label{bin2}
\end{eqnarray}
where:
\begin{equation}
(\vec{\xi ^{1}}\ast \vec{\xi ^{2}})_{r}=\sqrt{3}d_{rst}\xi _{s}^{1}\xi
_{t}^{2}  \label{with2}
\end{equation}
and:
\begin{equation}
Tr(H_{1}H_{2})=3\xi _{0}^{1}\xi _{0}^{2}+\frac{1}{2}\vec{\xi ^{1}}\cdot \vec{%
\xi ^{2}}  \label{with3}
\end{equation}
\end{itemize}

In particular, for traceless matrices:
\begin{equation}
Tr(H_{1}H_{2})=\frac{1}{2}\vec{\xi ^{1}}\cdot \vec{\xi ^{2}}  \label{par1}
\end{equation}
and
\begin{equation}
H_{1}\ast H_{2}=\frac{1}{6}(\vec{\xi ^{1}}\cdot \vec{\xi ^{2}})\lambda _{0}+%
\frac{1}{4\sqrt{3}}(\vec{\xi ^{1}}\ast \vec{\xi ^{2}})\cdot \vec{\lambda}
\label{par2}
\end{equation}
and, if $H_{3}$ is another traceless Hermitean matrix:
\begin{equation}
Tr\{(H_{1}\ast H_{2})H_{3}\}=\frac{1}{2}Tr\{(H_{1}H_{2}+H_{2}H_{1})H_{3}\}=%
\frac{1}{4\sqrt{3}}(\vec{\xi ^{1}}\ast \vec{\xi ^{2}})\cdot \vec{\xi ^{3}}
\label{par3}
\end{equation}
Quite obviously then, both:
\begin{equation}
\vec{\xi}\cdot \vec{\xi}=2Tr\{H(\xi )^{2}\}  \label{inv1}
\end{equation}
and:
\begin{equation}
(\vec{\xi}\ast \vec{\xi})\cdot \vec{\xi}=4\sqrt{3}Tr\{H(\xi )^{3}\}\equiv
\sqrt{3}d_{rst}\xi _{r}\xi _{s}\xi _{t}  \label{inv2}
\end{equation}
will be invariant under conjugation, and will provide us with a quadratic and a cubic invariant respectively
\cite{Ar,MM}. A third (linear) invariant
would be $Tr\{H\}$ whenever $H$ is not traceless. Notice that $(\vec{\xi}%
\ast \vec{\xi})\cdot \vec{\xi}$ is bounded both from above and from below,
and indeed it can be proved that:
\begin{equation}
-|\vec{\xi}|^{3}\leq (\vec{\xi}\ast \vec{\xi})\cdot \vec{\xi}\leq |\vec{\xi}%
|^{3}  \label{bound}
\end{equation}
where, in an obvious notation: $|\vec{\xi}|=:\sqrt{\vec{\xi}\cdot \vec{\xi}}$%
. Remember also that:
\begin{equation}
\det (H(\xi ))=\frac{1}{2\sqrt{3}}(\vec{\xi}\ast \vec{\xi})\cdot \vec{\xi}
\label{det}
\end{equation}

The eigenvalues of $H(\xi)$ (for $\xi_{0}=0$) can be conveniently expressed
as follows \cite{Ar}. For each given $\vec{\xi}$, we define first an angle $%
\phi$ in the range $[\pi/6,\pi/2]$ by:
\begin{equation}
(\vec{\xi}\ast\vec{\xi})\cdot \vec{\xi}=-|\vec{\xi}|^{3}\sin(3\phi)
\label{angle}
\end{equation}
Given the limitations on $(\vec{\xi}\ast\vec{\xi}) \cdot\vec{\xi}$ and the
range specified for $\phi$, the angle is uniquely determined by $\vec{\xi}$
(except for $\vec{\xi}=0$, of course) and is by construction an $SU(3)$
invariant. The eigenvalues of $H(\xi)$, written as $E_{a}(\vec{\xi})$, $%
a=1,2,3$, are given in nonincreasing order by:
\begin{eqnarray}
E_{1}(\vec{\xi}) &=& \frac{|\vec{\xi}|}{\sqrt{3}}\sin \phi  \label{ev1} \\
E_{2}(\vec{\xi}) &=& \frac{|\vec{\xi}|}{\sqrt{3}}\sin \left( \phi+ \frac{2\pi%
}{3} \right)  \label{ev2} \\
E_{3}(\vec{\xi}) &=& \frac{|\vec{\xi}|}{\sqrt{3}}\sin \left( \phi+ \frac{4\pi%
}{3} \right)  \label{ev3}
\end{eqnarray}
\begin{equation}
E_{1}(\vec{\xi})\geq E_{2}(\vec{\xi})\geq E_{3}(\vec{\xi}), \; E_{1}(\vec{\xi%
})+E_{2}(\vec{\xi})+E_{3}(\vec{\xi})=0  \label{ev}
\end{equation}
The successive energy differences have the simple forms:
\begin{equation}
E_{12}(\vec{\xi})=E_{1}(\vec{\xi})-E_{2}(\vec{\xi})=|\vec{\xi}|\sin \left(
\phi-\frac{\pi}{6}\right)  \label{d12}
\end{equation}
and:
\begin{equation}
E_{23}(\vec{\xi})=E_{2}(\vec{\xi})-E_{3}(\vec{\xi})=|\vec{\xi} | \cos \phi
\label{d23}
\end{equation}
We will sometimes write simply $E_{a},E_{ab}$, omitting explicit mention of $%
|\vec{\xi}|$.

It is easy to check that double degeneracies occur only for $\phi$ at the
extremes of the interval of definition, and precisely that:
\begin{equation}
\phi=\frac{\pi}{6}\Rightarrow E_{12}=0  \label{updeg}
\end{equation}
the upper double degeneracy,
\begin{equation}
\phi=\frac{\pi}{2}\Rightarrow E_{23}=0  \label{lowdeg}
\end{equation}
the lower double degeneracy, while for $\phi\in(\pi/6,\pi/2)$ the
Hamiltonian is nondegenerate, i.e.: $E_{12},E_{23} > 0$, while triple
degeneracy occurs only for $\vec{\xi}=0$.

Including dilations, the regions of double degeneracy are two distinct
nonoverlapping five-dimensional regions in $\br^{8}-\{\mathbf{0}\}$, each
one comprising a singular four-parameter family of directions. Denoting
these regions by $\Sigma _{12}$ and $\Sigma _{23}$ respectively:
\begin{equation}
\Sigma _{12}=\{\vec{\xi}\in \br^{8}-\{\mathbf{0}\}\;:\;(\vec{\xi}\ast \vec{%
\xi})\cdot \vec{\xi}=-|\vec{\xi}|^{3}\}  \label{s12}
\end{equation}
corresponding to $E_{12}=0,E_{23}>0$, and:
\begin{equation}
\Sigma _{23}=\{\vec{\xi}\in \br^{8}-\{\mathbf{0}\}\;:\;(\vec{\xi}\ast \vec{%
\xi})\cdot \vec{\xi}=+|\vec{\xi}|^{3}\}  \label{s23}
\end{equation}
corresponding to $E_{12}>0,E_{23}=0$.

We shall mostly deal with the generic, nondegenerate situation. Let us
denote the three orthonormal eigenvectors of $H(\xi )$ by $|a;\vec{\xi}%
\rangle $, $a=1,2,3$:
\begin{equation}
H(\xi )|a;\vec{\xi}\rangle =E_{a}(\vec{\xi})|a;\vec{\xi}\rangle \;,\;\langle
a;\vec{\xi}|b;\vec{\xi}\rangle =\delta _{ab}  \label{onev}
\end{equation}
The overall phases of these eigenvectors are at the moment free. The
standard orthonormal basis for the complex three-dimensional space $\bc^{3}$
is written simply as $|a\rangle $, with:
\begin{equation}
|1\rangle =\left[
\begin{array}{c}
1 \\
0 \\
0
\end{array}
\right] \,,\,|2\rangle =\left[
\begin{array}{c}
0 \\
1 \\
0
\end{array}
\right] \,,\,|3\rangle =\left[
\begin{array}{c}
0 \\
0 \\
1
\end{array}
\right] \,.  \label{c3}
\end{equation}
Then the matrix $A(\vec{\xi})$ defined by:
\begin{equation}
A_{ab}(\vec{\xi})=:\langle a|b;\vec{\xi}\rangle  \label{mat}
\end{equation}
is unitary, $A\in U(3)$, and relates the two bases:
\begin{equation}
|a;\vec{\xi}\rangle =A_{ba}(\vec{\xi})|b\rangle  \label{rel}
\end{equation}
(sums over repeated indices being understood). We will assume that the
phases of the three eigenvectors $|a;\vec{\xi}\rangle $ are adjusted in such
a way that actually $A(\vec{\xi })\in SU(3)$. This still leaves two free
phases in $A(\vec{\xi })$.

It is clear that conjugation of $H(\vec{\xi })$by $A(\vec{\xi })$ reduces it
to diagonal form. In this process $\vec{\xi}$ gets transformed to a $\vec{\xi%
}^{(0)}$ of which only the third and eighth components are nonzero and
moreover, in agreement with the ordering of the eigenvalues, the diagonal
entries of $H(\vec{\xi }^{(0)})$ are nonincreasing. We shall call
the resulting unique $\vec{\xi}^{(0)}$ the ``rest frame'' form \cite{MM} of $%
\vec{\xi}$. Therefore:
\begin{equation}
A(\vec{\xi})^{\dagger }H(\vec{\xi})A(\vec{\xi})=H(\vec{\xi}^{(0)})\;,\;\vec{%
\xi}^{(0)}=(0,0,\xi _{3}^{(0)},0,0,0,0,\xi _{8}^{(0)})  \label{rest}
\end{equation}
with: $\xi _{8}^{(0)}\geq \frac{\xi _{3}^{(0)}}{\sqrt{3}}\geq 0$ and
\begin{equation}
E_{1}=\frac{1}{2}(\xi _{3}^{(0)}+\frac{1}{\sqrt{3}}\xi _{8}^{(0)})\;,\;E_{2}=%
\frac{1}{2}(-\xi _{3}^{(0)}+\frac{1}{\sqrt{3}}\xi _{8}^{(0)})\;,\;E_{3}=-%
\frac{1}{\sqrt{3}}\xi _{8}^{(0)}  \label{en}
\end{equation}
Moreover:
\begin{equation}
\vec{\xi}^{(0)}\ast \vec{\xi}^{(0)}=(0,0,2\xi _{3}^{(0)}\xi
_{8}^{(0)},0,0,0,0,\xi _{3}^{(0)2}-\xi _{8}^{(0)2})  \label{f1}
\end{equation}
\begin{equation}
|\vec{\xi}|^{2}=\xi _{3}^{(0)2}+\xi _{8}^{(0)2}  \label{f2}
\end{equation}
and:
\begin{equation}
\vec{\xi}^{(0)}\ast \vec{\xi}^{(0)}\cdot \vec{\xi}^{(0)}=\xi _{8}^{(0)}(3\xi
_{3}^{(0)2}-\xi _{8}^{(0)2})  \label{f3}
\end{equation}
Notice that all this implies:
\begin{equation}
\xi _{3}^{(0)}=0\Leftrightarrow \vec{\xi}\ast \vec{\xi}\cdot \vec{\xi}=-|%
\vec{\xi}|^{3}\Leftrightarrow E_{12}=0  \label{f4}
\end{equation}
\begin{equation}
\xi _{3}^{(0)}=\sqrt{3}\xi _{8}^{(0)}\Leftrightarrow \vec{\xi}\ast \vec{\xi}%
\cdot \vec{\xi}=|\vec{\xi}|^{3}\Leftrightarrow E_{23}=0  \label{f5}
\end{equation}
It should be noted that, while $\vec{\xi}^{(0)}\ast \vec{\xi}^{(0)}\cdot
\vec{\lambda }$ is diagonal, in general $\vec{\xi}^{(0)}\ast \vec{\xi}^{(0)}$
is not in rest frame form.

For general $\vec{\xi}$, the phase freedom in the eigenvectors of $H(\xi)$,
even with the condition $A(\vec{\xi})\in SU(3)$ corresponds to the fact that
$\vec{\xi}^{(0)}$ has a nontrivial stability subgroup $U(1)\times
U(1)\subset SU(3)$, the torus subgroup. Therefore $A(\vec{\xi})$ remains
undefined up to such an element on the right. This is evident if we rewrite
the relation between $H(\xi)$ and $H(\xi^{(0)})$ as:
\begin{equation}
H(\xi)=A(\vec{\xi})H(\xi^{(0)})A(\vec{\xi})^{\dagger}  \label{restH}
\end{equation}
We shall often refer to this freedom on the right in the choice of $A(\vec{%
\xi})$ for each $\vec{\xi}\in \br^{8} -\{\mathbf{0}\}$.

\bigskip

\textbf{Remark.}\newline%
 If, for $\vec{\xi} \neq 0$ (i.e. on $\br
^{8}-\{{\bf{0}}\} $) we normalize $\vec{\xi}$  by requiring, e.g.: $\vec{\xi}\cdot \vec{\xi}=1$ or, equivalently, by
quotienting w.r.t. the dilations in $\br^{8}-\{0\}$,we obtain the sphere
 $S^{7}$ of normalized
vectors. By quotienting further w.r.t. the $U(1)$ isotropy group (the ''phase isotropy subgroup'') of the
vectors, we obtain eventually the projective plane $CP^{3}$.
All in all, we have the sequence of
fibrations:
\begin{equation}
\begin{array}{ccc}
\br^{8}-\{\mathbf{0}\} & \longleftarrow & \br^{\ast } \\
\downarrow & ~ & ~ \\
S^7 & \longleftarrow & U(1) \\
\downarrow & ~ & ~ \\
CP^{3} & ~ & ~
\end{array}
\end{equation}

This implies that a study of the geometry of the generic orbits may turn out to be relevant for the study of the
geometry of entangled states as well. This comes about because the manifold of pure states for a composite
system
of \ two two-level systems (sucs as photons or spin-$1/2$ particles) is
$\bc^{4}$, and the associated projective space is $CP^{3}$%
.Product pure states are given by the four dimensional $CP
^{1}\times CP^{1}$ and, in general, the pure states of a composite $%
2\times 2$ quantum system span $\ U(4)/U(3)\times U(1)\approx CP^{3}
\rightleftharpoons \br^{8}-\{\bf{0}\}/\br^{\ast }\times U(1)$, which is diffeomorphic to the
generic coadjoint orbits of $SU(3)$.

As stated in Sect.2, the coadjoint orbits are symplectic manifolds, the symplectic form being given by
Eq.$(19)$. $\mathbf{\omega }_{H}^{3}$ will be a volume-form and will define a differentiable measure on the
orbit.
When pulled back to $S^{7}$, and identifying for short $\mathbf{%
\omega }_{H}^{3}$ with its pull-back, $\mathbf{\omega }_{H}^{3}\wedge
d\theta $ will be a volume-form on $S^{7}$ and, similarly, $\mathbf{%
\omega }_{H}^{3}\wedge d\theta \wedge dr/r$ will be a volume-form on $%
\br^{8}-\{\mathbf{0}\}$, the latter providing  a (Bures \cite {Bur,Bur2}) measure on the space of density
matrices.

Finally, the orbits can also be equipped with a metric. Indeed, \ if $%
\{H_{i}\}$ is a basis in the Lie algebra $u(n)$ of $U(n)$ (e.g. the one
given in Eqs.$(5-7)$
 we can also define a left-invariant metric $\mathit{%
g}$ on $U(n)$ as:
\begin{equation}
\mathit{g}=\delta ^{ij}(TrH_{i}U^{\dagger }dU)\otimes (TrH_{j}U^{\dagger }dU) \label{metric1}
\end{equation}
and similarly for a right-invariant one, or a bi-invariant metric as:
\begin{equation}
\widetilde{\mathit{g}}=\text{ }Tr(U^{\dagger }dU\otimes U^{\dagger }dU)
\end{equation}
and on each orbit we have a metric like:
\begin{equation}
Tr([H,U^{\dagger }dU]\otimes \lbrack H,U^{\dagger }dU])
\end{equation}
or:
\begin{equation}
Tr([H,dUU^{\dagger }]\otimes \lbrack H,dUU^{\dagger }])
\end{equation}
according to which action has been chosen to define the quotient.

\bigskip

\section{Some Geometry of the Adjoint Orbits of $SU(3)$.}

\bigskip

As already stated in Sect.$2$, we will be interested in the orbits of $SU(3)$
in the space of the traceless Hermitean matrices or in those of $U(3)$ in
the space of all Hermitean matrices.

To make contact with the previous Section, we recall that each orbit is
characterized uniquely by the isotropy group of any one of its elements (the
isotropy groups for different elements being conjugate subgroups of $SU(3)$
or $U(3)$ as the case may be), and that the latter is generated by the
commutant in the way that has been discussed in Sect.$2$ and that will be
rephrased here in the specific context of $SU(3)$ (or $U(3)$).

If $H$ is generic, the isotropy group is generated by $H,H^{2}$ and $H^{3}$
as:
\begin{equation}
\{e^{is_{1}H},e^{is_{2}H^{2}},e^{is_{3}H^{3}}\}=U(1)\times U(1)\times U(1)
\label{gen}
\end{equation}
and the invariants characterizing the orbits in $U(3)$ will be: $%
Tr(H),Tr(H^{2})$ and $Tr(H^{3})$. If we restrict to traceless matrices we
are left with $Tr(H^{2})$ and $Tr(H^{3})$ and, according to the discussion
of Sect.$2$, we can replace $Tr(H^{3})$ with $Det(H)$. For more details see
\cite{Ki}. Also, the isotropy group in $SU(3)$ will be $U(1)\times U(1)$
(compare the discussion of the previous Section). Explicit expressions for
the invariants in terms of the vector $\vec{\xi}$ have been given before and
will not be reproduced here.

If $H$ is not generic and has a single doubly degenerate eigenvalue the
isotropy group will be $U(2)\subset SU(3)$ (or $U(2)\times U(1)$ in $U(3)$).
Finally, when $H$ has a triply degenerate eigenvalue the isotropy group will
become the whole of $SU(3)$ (or $U(3)$).

For any group $G$ and for any given $H\in \underline{G}$, the Lie algebra of
$G$, the orbit through $H$ will be diffeomorphic with the coset space $%
G/K_{H}$, with $K_{H}$ the isotropy group of $H$. We can also identify the
orbit by taking $H$ in its rest frame. With $H$ fixed in this manner, we can
write the symplectic structure on this coset space as:
\begin{equation}
\omega (H)=d\{TrHs^{-1}ds\}\;,\;s\in G  \label{omega1}
\end{equation}
(if $G=U(3)$ or $SU(3)$, $s^{-1}=s^{\dagger }$) or, using $%
ds^{-1}=-s^{-1}dss^{-1}$:
\begin{equation}
\omega (H)=-TrH\{s^{-1}ds\wedge s^{-1}ds\}  \label{omega2}
\end{equation}

Generic orbits will be defined in $\br^{8}$ by the algebraic equations:
\begin{eqnarray}
&&\vec{\xi}\cdot \vec{\xi}=\xi ^{2}=const.=c_{1}  \label{orbit1} \\
&&\vec{\xi}\ast \vec{\xi}\cdot \vec{\xi}=const.=c_{2}\neq \pm \xi ^{3}
\label{orbit2}
\end{eqnarray}
The intersection of these algebraic varieties yields the (generic)
symplectic orbit which is six-dimensional and is diffeomorphic with $%
SU(3)/U(1)\times U(1)\approx U(3)/U(1)\times U(1)\times U(1)\approx
CP^{3}$.

Four-dimensional exceptional orbits are defined by intersecting $\xi
^{2}=c_{1}$ with the subspaces $\Sigma _{12}$ or $\Sigma _{23}$ defined in
Sect.$3$. The intersections are diffeomorphic with $SU(3)/U(2)$.

\bigskip

\textbf{Remarks.}

\begin{enumerate}
\item  Assuming that $H$ be traceless does not seem to play a relevant role
in this context, and indeed the commutant of $H$ does not change if we add
or subtract from $H$ any multiple of the identity.

\item  Given a doubly degenerate eigenvalue for $H$, say:
\begin{equation}
H=\lambda \{|1\rangle \langle 1|+|2\rangle \langle 2|\}+\mu |3\rangle
\langle 3|\;,\;\lambda \neq \mu   \label{double}
\end{equation}
we may add $-\lambda \mathbf{I}$ to $H$ to get:
\begin{equation}
(\mu -\lambda )^{-1}(H-\lambda \mathbf{I)}=|3\rangle \langle 3|  \label{eq}
\end{equation}
Therefore the symplectic orbit through a pure state is diffeomorphic to the
symplectic orbit through a matrix with a doubly degenerate eigenvalue.

\item  The foliation of $U(3)$ defined by the level sets of the function:
\begin{equation}
iTr:u(3)\rightarrow \br  \label{fol}
\end{equation}
include orbits of the coadjoint action. By considering the intersection with
$(iTr)^{-1}(0)$ we find orbits of $SU(3)$ in the dual algebra of $su(3)$.

\item  Starting from the Hilbert space $\mathcal{H}$ we may consider the
projection: $\pi :\mathcal{H}\setminus \{0\}\rightarrow \mathcal{P(H)}$. In
Dirac's notation:
\begin{equation}
\pi :|\psi \rangle \mapsto \frac{|\psi \rangle \langle \psi |}{\langle \psi
|\psi \rangle }  \label{dirac}
\end{equation}
Notice that in this way \ $\mathcal{H}\setminus \{0\}$ becomes a principal $%
\bc$-bundle over $\mathcal{P(H)}$. Moreover:
\begin{equation}
\pi (|\psi \rangle +|\phi \rangle )\varpropto |\psi \rangle \langle \psi
|+|\phi \rangle \langle \phi |+(|\psi \rangle \langle \phi |+|\phi \rangle
\langle \psi |)  \label{pi1}
\end{equation}
and:
\begin{equation}
\pi (|\psi \rangle +i|\phi \rangle )\varpropto |\psi \rangle
\langle \psi |+|\phi \rangle \langle \phi |-i(|\psi \rangle
\langle \phi |-|\phi \rangle \langle \psi |)  \label{pi2}
\end{equation}
so under projection we generate both $|\psi \rangle \langle \phi |+|\phi \rangle \langle \psi |$ and $\ i(|\psi
\rangle \langle \phi |-|\phi \rangle \langle \psi |)$. Hence \textbf{all} Hermitean matrices can be written as
combinations \textbf{over the reals} of elements in $\mathcal{P(H)}$. Appropriate restrictions on the
coefficients will select the space of density matrices.
\end{enumerate}

\bigskip

\section{Geometric Phase Two-Forms and $U(1)$ Connections over Coadjoint
Orbits.}

\bigskip

After this general discussion of the structure of the space of all (pure and
mixed) states, we turn to the geometric phase problem for three-level
systems.

Following the Adiabatic Theorem of Quantum Mechanics we can imagine carrying
any one of the three nondegenerate eigenvectors $|a;\xi \rangle $ of the
(generic) Hamiltonian $H(\xi )$ along a closed path $\mathcal{C}\subset \br%
^{8}$, avoiding the origin and the subsets $\Sigma _{12\text{ }}$and $\Sigma
_{23}$, and we can then ask how much geometrical phase it has accumulated.
From a physical point of view we would consider circuits ``close'' to the
origin $\vec{\xi}=0$. Thus we are concerned with three different geometrical
phases $\phi ^{(a)}(\mathcal{C}),a=1,2,3$, each one associated with a given
eigenvector. Notice that in the case of a two-level system, when the
relevant group is $SU(2)$, one has also two in principle different
geometrical phases. However, as they differ by a constant ($4\pi $) and by a
sign, one is accustomed to speak of a single ``Berry phase'' in that
context. Here we will be forced to deal with three phases. We will discuss
in what follows some more general ``sum rule'' among them.

It is known that the above geometric phases can be computed as integrals
over any two-dimensional surface bounded by $\mathcal{C}$ in $\br%
^{8}-\{\{0\}\cup \Sigma_{12}\cup\Sigma_{23}\}$ of corresponding two-forms $%
V^{(a)}(\xi)$. Such forms are the curvature forms associated with the
corresponding ``Berry phase'' connection one-forms. The general expression
for these two-forms describing the ``flux'' of the geometric phase in
parameter space is \cite{Be}:
\begin{eqnarray}
V^{(a)}(\xi) &=& \mathrm{Im}\sum\limits_{b\neq a}\frac{\langle a;\xi|dH(\xi
)|b;\xi\rangle\wedge\langle b;\xi|dH(\xi)|a;\xi \rangle}{E_{ab}(\xi)^{2}}
\label{2form} \\
&=& \frac{1}{2}V_{rs}^{(a)}(\xi)d\xi_{r}\wedge d\xi_{s}  \notag
\end{eqnarray}
where:
\begin{equation}
V_{rs}^{(a)}(\xi)=\frac{1}{4}\mathrm{Im}\sum\limits_{b\neq a} \frac{\langle
a;\xi|\lambda_{r}|b;\xi \rangle\langle b;\xi|\lambda_{s}|a;\xi
\rangle-\{r\leftrightarrow s\}}{E_{ab}(\xi)^{2}}  \label{vcoeff}
\end{equation}

Before proceeding with the analysis of the $V^{(a)}$'s, let's pause a moment
and see how they arise in the geometric context of the (co)adjoint orbits of
$SU(3)$ in the space of Hermitean matrices.

Let then $H=H(\xi)$ be a Hermitean matrix (we will not need to impose here
conditions on its trace), and consider the eigenvalue problem:
\begin{equation}
H(\xi)|a;\xi\rangle =E_{a}(\xi)|a;\xi\rangle  \label{ep}
\end{equation}
along with the orbit of $H(\xi)$ under conjugation. By using, as a shorthand
for eq. (\ref{basis}), $H(\xi)=\frac{1}{2}\xi\cdot\lambda$ we may consider
the particular $\xi^{(0)}$ such that $H(\xi^{(0)})=:H_{0}$ is in diagonal
form, or in its rest frame. We will denote by $E_{a}=(E_{1},E_{2},E_{3})$
the eigenvalues.

If $|n\rangle $ denotes a normalized vector in the Hilbert space $\mathcal{H}
$ on which $H$ operates, a natural connection on $\mathcal{H}$ associated
with the projection $\pi :\mathcal{H}\setminus \{0\}\rightarrow \mathcal{P(H)%
}$ (or for the principal bundle $\bc^{\ast }\rightarrow \mathcal{H}\setminus
\{0\}\rightarrow \mathcal{P(H)}$) is defined by the parallel transport
condition:
\begin{equation}
\langle n|\overset{\cdot }{n}\rangle =0\Leftrightarrow \langle n|dn\rangle =0
\label{parallel}
\end{equation}
where $\overset{\cdot }{n}$ is evaluated along the ``transporting path''. If
the transport is unitary we may identify $|n\rangle $ with $|a;\xi \rangle $
with $\xi =\xi (t)$. We may also consider (cfr. Sect.$3$) the
``instantaneous'' matrix $A(\xi )$ such that
\begin{equation}
|a;\xi \rangle =A(\xi )|a\rangle  \label{a}
\end{equation}
where $\{|a\rangle \}$ is the standard basis (\ref{c3}). As discussed in
Sect.$3$, $A(\xi )$ brings by conjugation $H(\xi )$ to its diagonal form:
\begin{equation}
H(\xi )=A(\xi )H_{0}A(\xi )^{\dagger }  \label{diagH}
\end{equation}
and is determined only up to right multiplication by the isotropy group of $%
H_{0}$. Notice that this equation leads to:
\begin{equation}
dH=[dAA^{\dagger },H]=A[A^{\dagger }dA,H_{0}]A^{\dagger }  \label{dH}
\end{equation}

It is clear now that the parallel transport condition leads to: $\langle
a;\xi |(d/dt)|a;\xi \rangle=0$ and, equivalently, to:
\begin{equation}
\langle a;\xi|dA|a \rangle =0  \label{cond1}
\end{equation}
or to
\begin{equation}
\langle a|A^{\dagger}dA|a \rangle =0  \label{cond2}
\end{equation}
Written in this way, this exhibits $A^{\dagger}dA$ as a left-invariant
one-form, that can be written as: $A^{\dagger}dA=\theta^{r}\lambda_{r}$ in a
basis $\{\theta^{r}\}$ of left-invariant one-forms. Taking the exterior
differential and using again: $dA^{\dagger}=-A^{\dagger}dAA^{\dagger}$ we
may write the differential of $A^{\dagger}dA$ as $d(A^{\dagger}dA)=-A^{%
\dagger} dA\wedge A^{\dagger}dA$. To make connection with, e.g., the
discussion of Sects.$2$ and $4$, let us remark that the symplectic structure
on the orbit through $H_{0}$ (or $H$ for that matter) is given precisely by:
\begin{equation}
\omega_{H_{0}}=dTr\{H_{0}A^{\dagger}dA\}=-Tr\{H_{0}A^{\dagger}dA\wedge
A^{\dagger}dA\}  \label{omega0}
\end{equation}

By using the eigenvalue equation for $H$ we find:
\begin{equation}
dH|a;\xi \rangle=E_{a}d|a;\xi \rangle -H d|a;\xi \rangle  \label{h1}
\end{equation}
whence, taking scalar products:
\begin{equation}
\langle b;\xi|dH|a;\xi \rangle=(E_{a}-E_{b})\langle b;\xi|d|a;\xi \rangle
\label{h2}
\end{equation}
Therefore:
\begin{equation}
\frac{\langle b;\xi|A[A^{\dagger}dA,H_{0}]A^{\dagger}|a;\xi \rangle}{%
E_{a}-E_{b}} =\langle b;\xi|d|a;\xi \rangle=\frac{\langle
b|[A^{\dagger}dA,H_{0}]|a \rangle}{E_{a}-E_{b}}  \label{h3}
\end{equation}

Going back now to the form that has been given initially for the
geometric-phase (or curvature) two-forms $V^{(a)}$, we see that we can
rewrite them as the imaginary parts of:
\begin{equation}
\widetilde{V}^{(a)}=\sum\limits_{b\neq a}\frac{\langle a;\xi|[dAA^{\dagger},
H]|b;\xi\rangle\wedge\langle b;\xi|[dAA^{\dagger},H]|a;\xi \rangle}{%
E_{ab}{}^{2}}  \label{v1}
\end{equation}
Using explicitly: $[dAA^{\dagger},H]=dAA^{\dagger}H-HdAA^{\dagger}$, we find
next:
\begin{equation}
\widetilde{V}^{(a)}=\sum\limits_{b\neq a}\langle a;\xi|dAA^{\dagger}|b;\xi
\rangle\wedge\langle b;\xi|dAA^{\dagger}|a;\xi \rangle  \label{v2}
\end{equation}
i.e.:
\begin{equation}
\widetilde{V}^{(a)}=\langle a;\xi|dAA^{\dagger}\wedge dAA^{\dagger}|a;\xi
\rangle-\langle a;\xi|dAA^{\dagger}|a;\xi\rangle\langle a;\xi|
dAA^{\dagger}|a;\xi \rangle  \label{v3}
\end{equation}
and eventually:
\begin{equation}
\widetilde{V}^{(a)}=Tr\{|a;\xi\rangle\langle a;\xi|dAA^{\dagger}\wedge
dAA^{\dagger}\}=Tr\{|a;\xi\rangle\langle a;\xi|A^{\dagger}dA\wedge
A^{\dagger}dA\}  \label{v4}
\end{equation}
Thus:
\begin{equation}
\sum\limits_{a}E_{a}V^{(a)}=\mathrm{Im}Tr\{H_{0}A^{\dagger}dA\wedge
A^{\dagger}dA\}  \label{sum}
\end{equation}
gives the required relationship between the symplectic structure on the
coadjoint orbit through $H_{0}$ and the curvature forms of the Berry phase
connection. In a sense, this is also the generalized ``sum rule'' among the
three Berry phases that was mentioned at the beginning of this Section.

We can now relate the expressions of the coefficients of $V^{(a)}$ in a
general frame and in their rest frame by use of eq.ns (\ref{mat},\ref{transf}%
):
\begin{eqnarray}
V_{rs}^{(a)}(\vec{\xi}) &=&D_{ru}(A(\vec{\xi}))\,D_{sv}(A(\vec{\xi}%
))\,V_{uv}^{(a)}(\vec{\xi}^{(0)})  \label{vrest} \\
V_{rs}^{(a)}(\vec{\xi ^{(0)}}) &=&\frac{1}{4}\mathrm{Im}\sum\limits_{b\neq a}%
\frac{(\lambda _{u})_{ab}(\lambda _{v})_{ba}-(\lambda _{v})_{ab}(\lambda
_{u})_{ba}}{E_{ab}(\vec{\xi}^{(0)})^{2}}  \label{vrest0}
\end{eqnarray}
The ambiguity of $A(\vec{\xi})$ up to an $U(1)\times U(1)$ element on the
right does not affect this relation, since this is also the stability group
of $\vec{\xi}^{(0)}$. By a double application of (\ref{vrest0}), for any $%
A\in SU(3)$, we can relate $V_{rs}^{(a)}(D(A)\vec{\xi})$ to $V_{uv}^{(a)}(%
\vec{\xi})$. The argument rests on the fact that, if $\vec{\xi}^{\prime
}=D(A)\vec{\xi}$, then $A(\vec{\xi}^{\prime })$ which connects $\vec{\xi}%
^{\prime }$ to $\vec{\xi}^{(0)}$ can differ from $A(\vec{\xi})$ only by an
element of $U(1)\times U(1)$ on the right:
\begin{equation}
A(D(A)\vec{\xi})=A(\vec{\xi})\,L\;,\;\;L\in U(1)\times U(1)  \label{differ}
\end{equation}
Thus $L$ is a ``Wigner rotation''. Using $D(L)\vec{\xi}^{(0)}=\vec{\xi}%
^{(0)} $, we easily find:
\begin{equation}
\vec{\xi}^{\prime }=D(A)\vec{\xi}\;:\;V_{rs}^{(a)}(\vec{\xi}^{\prime
})=D_{ru}(A)D_{sv}(A)V_{uv}^{(a)}(\vec{\xi})\;.  \label{vtransf}
\end{equation}
Thus $V_{rs}^{(a)}(\vec{\xi})$ explicitly transforms in a covariant manner
and at $\vec{\xi}^{(0)}$ it is $U(1)\times U(1)$ invariant.

The $U(1)\times U(1)$ subgroup in $SU(3)$ is generated by $\lambda_3$ and $%
\lambda_8$. On a general $\vec{\xi}$ the effects are:
\begin{eqnarray}
D(e^{i\alpha \lambda_3}) \vec{\xi} &=& ( \xi_1 \cos(2\alpha) + \xi_2
\sin(2\alpha) , \xi_2 \cos(2\alpha) - \xi_1 \sin(2\alpha) , \xi_3 ,
\label{effect3} \\
&~& \xi_4 \cos\alpha + \xi_5 \sin\alpha , \xi_5 \cos\alpha - \xi_4
\sin\alpha ,  \notag \\
&~& \xi_6 \cos\alpha - \xi_7 \sin\alpha , \xi_7 \cos\alpha + \xi_6
\sin\alpha , \xi_8 )  \notag \\
D(e^{i\beta \lambda_8}) \vec{\xi} &=& ( \xi_1 , \xi_2 ,\xi_3 , \xi_4 \cos(%
\sqrt{3}\beta) + \xi_5 \sin(\sqrt{3}\beta) ,  \label{effect8} \\
&~& \xi_5 \cos(\sqrt{3}\beta) - \xi_4 \sin(\sqrt{3}\beta) , \xi_6 \cos(\sqrt{%
3}\beta) + \xi_7 \sin(\sqrt{3}\beta) ,  \notag \\
&~& \xi_7 \cos(\sqrt{3}\beta) - \xi_6 \sin(\sqrt{3}\beta) , \xi_8 )  \notag
\end{eqnarray}

Invariance under these transformations and antisymmetry in $r,s$ imply that
the only nonzero components of $V_{rs}^{(a)}(\vec{\xi}^{(0)})$ are possibly
those with $rs=12,21,45$, $54,67,76,38,83$. Detailed calculations using (\ref
{vrest0}) lead to the following independent nonvanishing elements:
\begin{equation}
\begin{array}{lll}
V_{12}^{(1)}(\vec{\xi}^{(0)})=\frac{1}{2E_{12}^{2}} & V_{45}^{(1)}(\vec{\xi}%
^{(0)})=\frac{1}{2E_{13}^{2}} & V_{67}^{(1)}(\vec{\xi}^{(0)})=0 \\
V_{12}^{(2)}(\vec{\xi}^{(0)})=-\frac{1}{2E_{12}^{2}} & V_{45}^{(2)}(\vec{\xi}%
^{(0)})=0 & V_{67}^{(2)}(\vec{\xi}^{(0)})=\frac{1}{2E_{23}^{2}} \\
V_{12}^{(3)}(\vec{\xi}^{(0)})=0 & V_{45}^{(3)}(\vec{\xi}^{(0)})=-\frac{1}{%
2E_{13}^{2}} & V_{67}^{(3)}(\vec{\xi}^{(0)})=-\frac{1}{2E_{23}^{2}}
\end{array}
\label{nonzero}
\end{equation}

We have now to use (\ref{vrest}) to transform back from the rest frame to
the general frame, knowing that we are dealing with invariant quantities
obeying eq. (\ref{vtransf}). This involves some $SU(3)$ irreducible tensor
analysis which we develop in the next section.

\bigskip

\section{$SU(3)$ tensor analysis of the geometric phase two-forms.}

\bigskip

The transformation law (\ref{vtransf}) shows that, for each level $a$, $%
V_{rs}^{(a)}(\vec{\xi})$ is a second rank tensor over the octet
representation of $SU(3)$. In Appendix A, we describe how such a tensor can
be decomposed into $SU(3)$-irreducible components belonging to the decouplet
$(\underline{10})$, antidecouplet $(\underline{10}^{\star })$ and octet $(%
\underline{8})$ UIR's of $SU(3)$, and then can be recovered by combinations
of these components. Based on the matrix elements $V_{rs}^{(a)}(\vec{\xi}%
^{(0)})$ given in (\ref{nonzero}), we must compute the nonvanishing
irreducible tensor components in the rest frame, for each $a$, and then use
eq. (\ref{vrest0}) to get the expression for general $\vec{\xi}$.

It is convenient to interchangeably use octet indices $r,s=1,\hdots,8$ and
contravariant and covariant $SU(3)$ tensor indices $b,c,d,e,f=1,2,3$. We
find that, in the rest frame and for each level $a=1,2,3$, the only
nonvanishing components of the $(\underline{10})$ and $(\underline{10}%
^\star) $ tensors $W^{(a)bcd}$, $\overline{W}^{(a)}_{bcd}$ are the $123$
components; these are the only $U(1)\times U(1)$ invariant ones. As for the
octet parts, in all cases $X^{(a)}_r$ turns out to be diagonal (and
traceless), so in the octet notation this means that only the $r=3$ and $r=8$
components remain. Again by (\ref{effect3},\ref{effect8}) these are $%
U(1)\times U(1)$ invariant.

We list in formula (\ref{irr}) the nonvanishing rest frame irreducible
tensor components for each $V_{rs}^{(a)}(\vec{\xi}^{(0)})$:
\begin{equation}
\begin{array}{ccccc}
~ & W^{(a)123} & \overline{W}_{123}^{(a)} & X_{3}^{(a)} & X_{8}^{(a)} \\
V_{rs}^{(1)}(\vec{\xi}^{(0)}) & i\left( \frac{1}{E_{13}^{2}}-\frac{1}{%
E_{12}^{2}}\right) & i\left( \frac{1}{E_{12}^{2}}-\frac{1}{E_{13}^{2}}\right)
& -\frac{1}{E_{12}^{2}}-\frac{1}{2E_{13}^{2}} & -\frac{\sqrt{3}}{2}\frac{1}{%
E_{13}^{2}} \\
V_{rs}^{(2)}(\vec{\xi}^{(0)}) & i\left( \frac{1}{E_{12}^{2}}-\frac{1}{%
E_{23}^{2}}\right) & i\left( \frac{1}{E_{23}^{2}}-\frac{1}{E_{12}^{2}}\right)
& \frac{1}{E_{12}^{2}}+\frac{1}{2E_{23}^{2}} & -\frac{\sqrt{3}}{2}\frac{1}{%
E_{23}^{2}} \\
V_{rs}^{(3)}(\vec{\xi}^{(0)}) & i\left( \frac{1}{E_{23}^{2}}-\frac{1}{%
E_{13}^{2}}\right) & i\left( \frac{1}{E_{13}^{2}}-\frac{1}{E_{23}^{2}}\right)
& \frac{1}{2E_{13}^{2}}-\frac{1}{2E_{23}^{2}} & \frac{\sqrt{3}}{2}\left(
\frac{1}{E_{13}^{2}}+\frac{1}{E_{23}^{2}}\right)
\end{array}
\label{irr}
\end{equation}

We note that the $(\underline{10})$ and $(\underline{10}^\star)$ components
are pure imaginary.

Now we have to obtain $V_{rs}^{(a)}(\vec{\xi})$ in a general frame,
expressing them as far as possible explicitly in terms of $\vec{\xi}$. First
we consider the contributions from the octet components $X_{r}^{(a)}(\vec{\xi%
}^{(0)})$. To begin with, for each $a$ decompose $X_{r}^{(a)}(\vec{\xi}%
^{(0)})$ as a linear combination of the available rest frame octet vectors $%
\vec{\xi}^{(0)}$ and $\vec{\eta}^{(0)}$, where $\vec{\eta}^{(0)}=%
\overrightarrow{\xi }^{(0)}\ast \overrightarrow{\xi }^{(0)}$. After some
algebra we find:
\begin{eqnarray}
X_{r}^{(a)}(\vec{\xi}^{(0)}) &=&\left[ \xi _{3}^{(0)}\left( \xi
_{3}^{(0)2}-3\xi _{8}^{(0)2}\right) \right] ^{-1}\left( \lambda ^{(a)}\xi
_{r}^{(0)}+\mu ^{(a)}\eta _{r}^{(0)}\right)  \label{oct} \\
\lambda ^{(1)} &=&\frac{\sqrt{3}\eta _{3}^{(0)}-\eta _{8}^{(0)}}{2E_{13}^{2}}%
-\frac{\eta _{8}^{(0)}}{E_{12}^{2}}  \notag \\
\lambda ^{(2)} &=&\frac{\sqrt{3}\eta _{3}^{(0)}+\eta _{8}^{(0)}}{2E_{23}^{2}}%
+\frac{\eta _{8}^{(0)}}{E_{12}^{2}}  \notag \\
\lambda ^{(3)} &=&\frac{\eta _{8}^{(0)}-\sqrt{3}\eta _{3}^{(0)}}{2E_{13}^{2}}%
-\frac{\eta _{8}^{(0)}+\sqrt{3}\eta _{3}^{(0)}}{2E_{23}^{2}}  \notag \\
\mu ^{(1)} &=&\frac{\xi _{8}^{(0)}}{E_{12}^{2}}+\frac{\xi _{8}^{(0)}-\sqrt{3}%
\xi _{3}^{(0)}}{2E_{13}^{2}}  \notag \\
\mu ^{(2)} &=&-\frac{\xi _{8}^{(0)}}{E_{12}^{2}}-\frac{\xi _{8}^{(0)}+\sqrt{3%
}\xi _{3}^{(0)}}{2E_{23}^{2}}  \notag \\
\mu ^{(3)} &=&\frac{\sqrt{3}\xi _{3}^{(0)}-\xi _{8}^{(0)}}{2E_{13}^{2}}+%
\frac{\sqrt{3}\xi _{3}^{(0)}+\xi _{8}^{(0)}}{2E_{23}^{2}}  \notag
\end{eqnarray}
Apart from the expected squares of energy denominators, combinations of the
components of $\vec{\xi}^{(0)}$ and $\vec{\eta}^{(0)}$ appear. They are
expressible in terms of $E_{ab}$ using:
\begin{eqnarray}
&~&\xi _{3}^{(0)}=E_{12}\;,\;\xi _{8}^{(0)}=\frac{E_{13}+E_{23}}{2\sqrt{3}}%
\;;  \label{exp} \\
&~&\eta _{3}^{(0)}=\frac{E_{12}(E_{13}+E_{23})}{\sqrt{3}}\;,\;\eta
_{8}^{(0)}=E_{12}^{2}-\frac{(E_{13}+E_{23})^{2}}{12}\;;  \notag \\
&~&\xi _{3}^{(0)}\left( \xi _{3}^{(0)2}-3\xi _{8}^{(0)2}\right)
=-4E_{12}E_{13}E_{23}  \notag
\end{eqnarray}
Combining the appropriate parts of eq.ns (\ref{project},\ref{comp8},\ref
{total}), we see that in a general frame the octet part of $V_{rs}^{(a)}(%
\vec{\xi})$ is given by
\begin{equation}
-\frac{1}{3}f_{rst}X_{t}^{(a)}(\vec{\xi})=\frac{1}{12E_{12}E_{13}E_{23}}%
f_{rst}\left( \lambda ^{(a)}\xi _{t}+\mu ^{(a)}\eta _{t}\right)
\label{part8}
\end{equation}
Actually the combinations of $\vec{\xi}$ and $\vec{\eta}=\overrightarrow{\xi
}\ast $ $\overrightarrow{\xi }$ occurring here become singular near the
regions of double degeneracy, $E_{12}\rightarrow 0$ or $E_{23}\rightarrow 0$%
. We examine these details in Sec.$7$.

The $(\underline{10})$ and $(\underline{10}^{\star })$ contributions to $%
V_{rs}^{(a)}(\vec{\xi})$ are more subtle in structure. We have seen that in
the rest frame the only nonzero components of $W^{(a)bcd}$ and $\overline{W}%
_{bcd}^{(a)}$ are the ones with $bcd=123$ (or permutations thereof). Let us
define a numerical tensor of type $(\underline{10})$ in the rest frame by
\begin{equation}
\delta ^{abc}=\left\{
\begin{array}{ll}
1 & \mbox{if $abc=$ any permutation of $123$} \\
0 & \mbox{otherwise}
\end{array}
\right.  \label{dup}
\end{equation}
Thus the only independent nonzero component of $\delta ^{abc}$ is the one
invariant under $U(1)\times U(1)$, the stability group of $\vec{\xi}^{(0)}$.
Then in the general frame we define the $(\underline{10})$ tensor
\begin{equation}
\Delta ^{abc}(\vec{\xi})=A_{d}^{a}(\vec{\xi})A_{e}^{b}(\vec{\xi})A_{f}^{c}(%
\vec{\xi})\delta ^{def}  \label{deltaup}
\end{equation}
The ambiguity in $A(\vec{\xi})$ up to a $U(1)\times U(1)$ element on the
right leaves $\delta ^{def}$ unaffected. Therefore it is consistent to
maintain that the quantities $\Delta ^{abc}(\vec{\xi})$ are the components
of a decouplet or $(\underline{10})$ tensor under $\vec{\xi}\rightarrow \vec{%
\xi}^{\prime }=A\vec{\xi}$ and, as written, are well defined functions of
the octet vector $\vec{\xi}$, in spite of the practical difficulty in
developing the expression (\ref{deltaup}) further. In a similar way, we set
up a rest frame numerical $(\underline{10}^{\star })$ tensor $\overline{%
\delta }_{abc}$, and then transport it to a general frame to get a $(%
\underline{10}^{\star })$ tensor $\overline{\Delta }_{abc}(\vec{\xi})$:
\begin{equation}
\overline{\delta }_{abc}=\left\{
\begin{array}{ll}
1 & \mbox{if $abc=$ any permutation of $123$} \\
0 & \mbox{otherwise}
\end{array}
\right.  \label{ddown}
\end{equation}
\begin{equation}
\overline{\Delta }_{abc}(\vec{\xi})=A_{a}^{d}(\vec{\xi})A_{b}^{e}(\vec{\xi}%
)A_{c}^{f}(\vec{\xi})\delta _{def}  \label{deltadown}
\end{equation}
Picking up the terms in eq. (\ref{total}) involving $W$ and $\overline{W}$,
and using (\ref{irr}), we get the remaining decouplet parts of $%
V_{~de}^{(a)bc}(\vec{\xi})$, expressed in tensor index notation:
\begin{equation}
\frac{iv^{(a)}}{6}\left( \epsilon _{def}\text{ }\Delta ^{bcf}(\vec{\xi}%
)-\epsilon ^{bcf}\text{ }\overline{\Delta }_{def}(\vec{\xi})\right)
\label{part10}
\end{equation}
with:
\begin{equation}
v^{(1)}=\frac{1}{E_{13}^{2}}-\frac{1}{E_{12}^{2}}\;,\;v^{(2)}=\frac{1}{%
E_{12}^{2}}-\frac{1}{E_{23}^{2}}\;,\;v^{(3)}=\frac{1}{E_{23}^{2}}-\frac{1}{%
E_{13}^{2}}\;.  \label{v123}
\end{equation}
The complete two-forms $V_{rs}^{(a)}(\vec{\xi})$ are obtained by putting
together the expressions in eq.ns (\ref{part8},\ref{part10}).

\bigskip

\section{Behaviour near double degeneracy.}

\bigskip

>From the description of the eigenvalue spectrum of $H(\vec{\xi})$ in Sect.$3$%
, it is clear that the energy differences obey
\begin{eqnarray}
&&E_{12}\,,\,E_{23}\geq 0\;,  \label{diff} \\
&&E_{13}=E_{12}+E_{23}\geq E_{12}\,,\,E_{23}\;.  \notag
\end{eqnarray}
The vanishing of $E_{13}$ occurs only at the point of triple degeneracy $%
\vec{\xi}=0$, the origin in $\br^{8}$. Away from this point, $E_{13}$ is
always strictly positive. However, as pointed out in Sect.$3$, there are
particular directions in $\br^{8}$ along which either $E_{12}$ or $E_{23}$
vanishes, signalling a double degeneracy. These are the two separate
four-parameter sets of directions comprising the regions $\Sigma _{12}$, $%
\Sigma _{23}$ defined in eq.ns (\ref{s12},\ref{s23}). As explained in
Appendix A, these directions map out the two four-dimensional regions in $%
S^{7}$ which correspond to the two singular orbits in the Lie algebra $su(3)$
of $SU(3)$, each realizing the coset space $SU(3)/U(2)$. We can exhibit the
structures of the two basic energy differences $E_{12}$, $E_{23}$ in the
vicinity of these regions, expressing them in terms of the two $SU(3)$
invariants $\vec{\xi}^{2}$ and $\vec{\xi}\cdot \vec{\xi}\ast \vec{\xi}$.
Starting with eq.ns (\ref{angle}) and (\ref{ev}) and expanding them near $%
\phi =\pi /6$ and $\phi =\pi /2$ respectively, we find:
\begin{eqnarray}
\phi \approx \frac{\pi }{6},\;\vec{\xi}\mbox{ near }\Sigma _{12} &:&E_{12}(%
\vec{\xi})\approx \frac{\sqrt{2}}{3}\,\frac{(\xi ^{3}+\vec{\xi}\cdot \vec{\xi%
}\ast \vec{\xi})^{1/2}}{\xi ^{1/2}}  \label{pi6} \\
\phi \approx \frac{\pi }{2},\;\vec{\xi}\mbox{ near }\Sigma _{23} &:&E_{23}(%
\vec{\xi})\approx \frac{\sqrt{2}}{3}\,\frac{(\xi ^{3}-\vec{\xi}\cdot \vec{\xi%
}\ast \vec{\xi})^{1/2}}{\xi ^{1/2}}  \label{pi}
\end{eqnarray}

Now we analyze in more detail the situation near, say, the upper double
degeneracy, when $\vec{\xi}$ lies close to $\Sigma _{12}$. First we deal
with the octet or $X^{(a)}$ contribution to $V_{rs}^{(a)}(\vec{\xi})$ in
this limit, and later look at the $(\underline{10})$ and $(\underline{10}%
^{\star })$ contributions. The aim is to trace how the two-level monopole
structure is recovered from the present three-level formalism. It is
adequate to work in the rest frame $\vec{\xi}=\vec{\xi}^{(0)}$ and use eq. (%
\ref{oct}), since the transition to a general frame via $SU(3)$ cannot
introduce any singular factors. Let us in the following denote the small
energy difference $E_{12}$ by $\epsilon $. The prefactor in eq. (\ref{oct})
introduces an explicit $1/\epsilon $ factor in the expression for $X^{(a)}$:
\begin{equation}
\left[ \xi _{3}^{(0)}\left( \xi _{3}^{(0)2}-3\xi _{8}^{(0)2}\right) \right]
^{-1}=-\frac{1}{4\epsilon E_{13}E_{23}}  \label{pre}
\end{equation}
This must be balanced by matching powers of $\epsilon $ from the remaining
parts of eq. (\ref{oct}), since we know from the general formula (\ref{2form}%
) that nothing more singular than $1/\epsilon ^{2}$ can appear in the
two-form $V^{(a)}$. This is indeed borne out by detailed calculations. We
find, remembering that $E_{12}=\epsilon $ is the small parameter, that the $%
3\times 3$ hermitian matrices $X^{(a)}(\xi ^{(0)})$ have the following
structures:
\begin{eqnarray}
X^{(1)}(\xi ^{(0)}) &=&-\frac{1}{4E_{13}E_{23}}\left[ 2\left( 2\frac{%
E_{13}^{2}}{\epsilon ^{2}}-2\frac{E_{13}}{\epsilon }+1-\frac{\epsilon }{%
E_{13}}\right) \lambda _{3}\right.  \label{chi1} \\
&+&\left. 2\sqrt{3}\left( 1-\frac{\epsilon }{E_{13}}\right) \lambda _{8}%
\right] =-\frac{\lambda _{3}}{\epsilon ^{2}}+%
\mbox{\small{nonsingular
terms}}  \notag \\
X^{(2)}(\xi ^{(0)}) &=&-\frac{1}{4E_{13}E_{23}}\left[ -2\left( 2\frac{%
E_{23}^{2}}{\epsilon ^{2}}+2\frac{E_{23}}{\epsilon }+1+\frac{\epsilon }{%
E_{23}}\right) \lambda _{3}\right.  \label{chi2} \\
&+&\left. 2\sqrt{3}\left( 1+\frac{\epsilon }{E_{23}}\right) \lambda _{8}%
\right] =\frac{\lambda _{3}}{\epsilon ^{2}}+\mbox{\small{nonsingular
terms}}  \notag \\
X^{(3)}(\xi ^{(0)}) &=&\mbox{\small{ only nonsingular terms}}  \label{chi3}
\end{eqnarray}
Therefore from eq. (\ref{part8}) the leading nonvanishing octet
contributions to $V_{rs}^{(a)}(\vec{\xi}^{(0)})$, apart from antisymmetry in
$r$ and $s$, are then:
\begin{equation}
\begin{array}{lll}
V_{12}^{(1)}(\vec{\xi}^{(0)})=\frac{1}{3\epsilon ^{2}}\;, & V_{45}^{(1)}(%
\vec{\xi}^{(0)})=\frac{1}{6\epsilon ^{2}}\;, & V_{67}^{(1)}(\vec{\xi}%
^{(0)})=-\frac{1}{6\epsilon ^{2}}\;, \\
V_{12}^{(2)}(\vec{\xi}^{(0)})=-\frac{1}{3\epsilon ^{2}}\;, & V_{45}^{(2)}(%
\vec{\xi}^{(0)})=-\frac{1}{6\epsilon ^{2}}\;, & V_{67}^{(2)}(\vec{\xi}%
^{(0)})=\frac{1}{6\epsilon ^{2}}\;, \\
V_{rs}^{(3)}(\vec{\xi}^{(0)})=\mbox{\small{nonsingular}}\;. &  &
\end{array}
\label{comb8}
\end{equation}
On the other hand, from (\ref{irr}) and (\ref{v123}) we find that the
leading nonvanishing $(\underline{10})$ and $(\underline{10}^{\star })$
contributions to $V_{rs}^{(a)}(\vec{\xi}^{(0)})$, again apart from
antisymmetry in $r$ and $s$, are as follows:
\begin{equation}
\begin{array}{lll}
V_{12}^{(1)}(\vec{\xi}^{(0)})=\frac{1}{6\epsilon ^{2}}\;, & V_{45}^{(1)}(%
\vec{\xi}^{(0)})=-\frac{1}{6\epsilon ^{2}}\;, & V_{67}^{(1)}(\vec{\xi}%
^{(0)})=\frac{1}{6\epsilon ^{2}}\;, \\
V_{12}^{(2)}(\vec{\xi}^{(0)})=-\frac{1}{6\epsilon ^{2}}\;, & V_{45}^{(2)}(%
\vec{\xi}^{(0)})=\frac{1}{6\epsilon ^{2}}\;, & V_{67}^{(2)}(\vec{\xi}%
^{(0)})=-\frac{1}{6\epsilon ^{2}}\;, \\
V_{rs}^{(3)}(\vec{\xi}^{(0)})=\mbox{\small{nonsingular}}\;. &  &
\end{array}
\label{comb10}
\end{equation}
Adding the two sets of contributions we see that for $\vec{\xi}$ near $%
\Sigma _{12}$, in the rest frame, the surviving singular terms in the
two-forms $V_{rs}^{(a)}(\vec{\xi}^{(0)})$ are very few:
\begin{equation}
V_{12}^{(1)}(\vec{\xi}^{(0)})=-V_{12}^{(2)}(\vec{\xi}^{(0)})=\frac{1}{%
2\epsilon ^{2}}  \label{sing}
\end{equation}
We can see how the different irreducible $SU(3)$ tensor components combine
in just the right manner to reproduce the three-dimensional magnetic
monopole type of singularity in the two-forms $V_{rs}^{(a)}(\vec{\xi}^{(0)})$
when we are near the point of double-degeneracy. Limiting ourself to $SU(2)$
transformations in the $12$ subspace in $SU(3)$, it is evident that the
terms $V_{12}^{(1,2)}(\vec{\xi}^{(0)})$ of eq. (\ref{sing}) give rise to the
familiar three-dimensional monopole field, expressed however in
eight-dimensional space using the relations in (\ref{pi2}). Thus they are
singular not only at a single point but all over the region $\Sigma
_{12}\subset \br^{8}-\{\vec{0}\}$.

\bigskip

\section{Concluding remarks.}

\bigskip

We have presented a study of two aspects of multilevel quantum systems,
unified by the geometrical features of coadjoint orbits in the Lie algebra
of the unitary groups $U(n)$, $SU(n)$. These are the properties of unitary
von Neumann evolution of general pure or mixed states for such systems; and
in the $n=3$ case the detailed structures of the geometric phases in the
neighborhood of degeneracies.\newline
We have examined the structures of the two-forms or covariant antisymmetric
tensor fields in parameter space, whose fluxes through any closed circuit
give the corresponding quantum adiabatic geometric phases when one is in the
vicinity of a point of three-level degeneracy. In comparison to the
three-dimensional monopole singularity in the two-level case, here one has a
much richer structure with many new features. To begin with, the parameter
space is eight-dimensional. In the spirit of the von-Neumann Wigner theorem,
double degeneracies occur along two four-parameter sets of singular rays, $%
\Sigma _{12}$ and $\Sigma _{23}$, in parameter space; while the triple
degeneracy occurs just at a single point $\vec{\xi}=\vec{0}$. Beyond this
there is a rich $SU(3)$ tensor structure in the relevant two-forms. Whereas
for the two-level case the result was a simple radial ``vector'' field, here
we have three independent irreducible tensor contributions belonging to the
tensor types $(\underline{8})$, $(\underline{10})$ and $(\underline{10}%
^{\star })$. Overall reality makes the last two complex conjugates of one
another, but intrinsically they should be considered as independent of one
another and of the $(\underline{8})$ contribution.

The complexity of the expressions we have obtained is unavoidable, and
automatically belongs to the next most interesting case from the point of
view of general quantum mechanics after the two-level degeneracy. We have
shown that they contain within them, embedded in intricate ways, the
monopole of the double degeneracy problem when one is near one of the
regions $\Sigma _{12}$ and $\Sigma _{23}$, in parameter space. However we
emphasize that in the complete eight-dimensional picture these are far from
being point singularities. We can explain the complexity of the present
expressions by saying that they have to contain the earlier monopole results
in certain limiting situations, and then go beyond them to handle a triple
degeneracy.

\bigskip

\bigskip \textbf{\LARGE Appendix A} 

\bigskip We collect here some basic information about the defining and
adjoint representations of $SU(3)$. The group $SU(3)$ is defined as follows:
\begin{equation}
SU(3)=\{A=3\times 3\;complex\;matrix\;:\;A^{\dagger }A=\mathbf{I}%
\;,\;detA=1\}\;.  \label{su3}
\end{equation}
The eight independent hermitian traceless generators are the Gell'mann $%
\lambda $-matrices:
\begin{eqnarray}
\lambda _{1}=\left[
\begin{array}{ccc}
0 & 1 & 0 \\
1 & 0 & 0 \\
0 & 0 & 0
\end{array}
\right] &,&\lambda _{2}=\left[
\begin{array}{ccc}
0 & -i & 0 \\
i & 0 & 0 \\
0 & 0 & 0
\end{array}
\right]  \label{gell} \\
\lambda _{3}=\left[
\begin{array}{ccc}
1 & 0 & 0 \\
0 & -1 & 0 \\
0 & 0 & 0
\end{array}
\right] &,&\lambda _{4}=\left[
\begin{array}{ccc}
0 & 0 & 1 \\
0 & 0 & 0 \\
1 & 0 & 0
\end{array}
\right]  \notag \\
\lambda _{5}=\left[
\begin{array}{ccc}
0 & 0 & -i \\
0 & 0 & 0 \\
i & 0 & 0
\end{array}
\right] &,&\lambda _{6}=\left[
\begin{array}{ccc}
0 & 0 & 0 \\
0 & 0 & 1 \\
0 & 1 & 0
\end{array}
\right]  \notag \\
\lambda _{7}=\left[
\begin{array}{ccc}
0 & 0 & 0 \\
0 & 0 & -i \\
0 & i & 0
\end{array}
\right] &,&\lambda _{8}=\frac{1}{\sqrt{3}}\left[
\begin{array}{ccc}
1 & 0 & 0 \\
0 & 1 & 0 \\
0 & 0 & -2
\end{array}
\right]  \notag
\end{eqnarray}
that obey the following rule:
\begin{equation}
Tr\{\lambda _{r}\lambda _{s}\}=2\delta _{rs}  \label{r1}
\end{equation}
Their commutation and anticommutation relations involve the completely
antisymmetric structure constants $f_{rst}$ and the completely symmetric $d$%
-symbols $d_{rst}$:
\begin{equation}
\lbrack \lambda _{r},\lambda _{s}]=2if_{rst}\lambda _{t}  \label{r2}
\end{equation}
and:
\begin{equation}
\{\lambda _{r},\lambda _{s}\}=\frac{4}{3}\delta _{rs}+2d_{rst}\lambda _{t}
\label{r3}
\end{equation}
($r,s,t=1,...,8$) where the numerical values of the independent $f_{rst}$
and $d_{rst}$ are
\begin{equation}
\begin{array}{l}
f_{123}=1\;,\;f_{458}=f_{678}=\frac{\sqrt{3}}{2}\;, \\
f_{147}=f_{246}=f_{257}=f_{345}=f_{516}=f_{637}=\frac{1}{2}\;.
\end{array}
\label{f}
\end{equation}
and
\begin{equation}
\begin{array}{l}
d_{118}=d_{228}=d_{338}=-d_{888}=\frac{1}{\sqrt{3}}\;,%
\;d_{448}=d_{558}=d_{668}=d_{778}=-\frac{1}{2\sqrt{3}}\;, \\
d_{146}=d_{157}=-d_{247}=d_{256}=d_{344}=d_{355}=-d_{366}=-d_{377}=\frac{1}{2%
}\;.
\end{array}
\label{d}
\end{equation}

Under conjugation by any $A\in SU(3)$ the $\lambda $'s go into real
orthogonal linear combinations of themselves:
\begin{equation}
\begin{array}{l}
A^{\dagger }\lambda _{r}A=D_{rs}(A)\lambda _{s}\;, \\
D_{rs}(A)=\frac{1}{2}Tr\{\lambda _{r}A\lambda _{s}A^{\dagger }\}\;; \\
D(A)^{T}D(A)=\mathbf{I}\;; \\
D(A^{\prime })D(A)=D(A^{\prime }A)\;.
\end{array}
\label{transf}
\end{equation}
These matrices constitute the octet or adjoint representation of $SU(3)$,
actually a faithful representation of the quotient $SU(3)/\mathbb{Z}_{3}$.
They describe the action of $SU(3)$ on a general eight component real octet
vector, as a small subgroup of the full twenty eight parameter group $SO(8)$%
. Given $\vec{\xi ^{1}},\vec{\xi ^{2}}\in \br^{8}$, we can form the $SU(3)$
(and $SO(8)$) invariant inner product:
\begin{equation}
\vec{\xi ^{1}}\cdot \vec{\xi ^{2}}=\vec{\xi ^{1}}_{r}\vec{\xi ^{2}}_{r}\;,
\label{asca}
\end{equation}
the antisymmetric octet vector $\vec{\xi ^{1}}\wedge \vec{\xi ^{2}}$ using
the structure constants:
\begin{equation}
(\vec{\xi ^{1}}\wedge \vec{\xi ^{2}})_{r}=-\frac{1}{2}f_{rst}\xi _{s}^{1}\xi
_{t}^{2}\;,  \label{awedge}
\end{equation}
as well as a symmetric octet vector $\vec{\xi ^{1}}\ast \vec{\xi ^{2}}$
using the $d$-symbols:
\begin{equation}
(\vec{\xi ^{1}}\ast \vec{\xi ^{2}})_{r}=\sqrt{3}d_{rst}\xi _{s}^{1}\xi
_{t}^{2}\;.  \label{astar}
\end{equation}
The latter two definitions make sense only with respect to $SU(3)$ and not $%
SO(8)$.

>From a given $\vec{\xi}\in \br^{8}$ one can construct the $SU(3)$ invariant $%
|\xi |^{2}=\vec{\xi}\cdot \vec{\xi}$ and $\vec{\xi}\cdot \vec{\xi}\ast \vec{%
\xi}$. The action of $SU(3)$ on $S^{7}$ is intricate. Denote by $\vec{n}$ a
unit octet vector, so $|\vec{n}|^{2}=1$. The remaining $SU(3)$ invariant is
the cubic $\vec{n}\cdot \vec{n}\ast \vec{n}$ and it obeys
\begin{equation}
-1\leq \vec{n}\cdot \vec{n}\ast \vec{n}\leq 1\;.  \label{limit}
\end{equation}
As long as this invariant is in the open interval $(-1,1)$, the stability
group of $\vec{n}$ is conjugate to $U(1)\times U(1)\subset SU(3)$; hence the
orbit of $\vec{n}$ under $SU(3)$ action, $D(A)\vec{n}$ $\forall A$, is
six-dimensional and realizes the coset space $SU(3)/U(1)\times U(1)$. The
six dimensions of this orbit together with the variable parameter $\vec{n}%
\cdot \vec{n}\ast \vec{n}$ account for the seven dimensions of $S^{7}$ at generic points. At the endpoints of
the interval in (\ref{limit}), the corresponding $\vec{n}\in S^{7}$ have four-dimensional stability groups,
conjugate to $U(2)\subset SU(3)$. Correspondingly we have two very special and singular four-dimensional orbits
of such vectors in $S^{7}$. In the notation of eq.ns (\ref{s12},\ref{s23}), they are given by
\begin{eqnarray}
\Sigma _{12}\cap S^{7} &=&\{\vec{n}\in S^{7}\;:\;\vec{n}\cdot \vec{n}\ast
\vec{n}=-1\}\;,  \label{ss12} \\
\Sigma _{23}\cap S^{7} &=&\{\vec{n}\in S^{7}\;:\;\vec{n}\cdot \vec{n}\ast
\vec{n}=+1\}\;.  \label{ss23}
\end{eqnarray}

\bigskip

\textbf{\LARGE Appendix B}

\bigskip

We develop here the algebraic tools to deal with the irreducible
tensor components of a second rank antisymmetric $SU(3)$ tensor
over its octet representation.

For an octet vector we can pass between $X_{r}$ and its tensor components by
using the $\lambda $ matrices:
\begin{eqnarray}
X_{b}^{a} &=&X_{r}(\lambda _{r})_{ab}\;,\;X_{a}^{a}=0  \label{xab} \\
X_{r} &=&\frac{1}{2}Tr\{X\lambda _{r}\}=\frac{1}{2}X_{b}^{a}(\lambda
_{r})_{ab}\;;  \notag \\
X_{r}^{\star } &=&X_{r}\Leftrightarrow (X_{b}^{a})^{\star }=X_{a}^{b}\;.
\notag
\end{eqnarray}
For ease in writing, the matrix indices on the $\lambda $'s are all given as
subscripts.

Now let $T_{rs}=T_{rs}^{\star }=-T_{sr}$ be a real antisymmetric second rank
tensor over the octet representation. Generalizing (\ref{xab}) we define its
tensor components $T_{cd}^{ab}$ as follows:
\begin{eqnarray}
&&T_{cd}^{ab}=(T_{ab}^{cd})^{\star }=-T_{dc}^{ba}=(\lambda
_{r})_{ac}(\lambda _{s})_{bd}T_{rs}\;,  \label{tensor} \\
&&T_{ac}^{ab}=T_{ba}^{ab}=0\;,  \notag \\
&&T_{rs}=\frac{1}{4}(\lambda _{r})_{ca}(\lambda _{s})_{db}T_{cd}^{ab}\;.
\notag
\end{eqnarray}
Such a tensor has 28 independent real components. Now, the antisymmetric
part of the direct product of two octet representations of $SU(3)$ contains,
upon reduction, the (complex) decouplet $\underline{10}$, the (conjugate)
anti-decouplet $\underline{10}^{\star }$ and the (real) octet $\underline{8}$%
, once each. The counting agrees since reality of $T_{rs}$ implies that the $%
\underline{10}^{\star }$ components are complex conjugates of the $%
\underline{10}$ components. We project out the $\underline{10}$, $\underline{%
10}^{\star }$ and $\underline{8}$ components respectively, of $T_{rs}$ in
the following way:
\begin{eqnarray}
W^{abc} &=&\epsilon ^{ade}T_{de}^{bc}+\epsilon ^{bde}T_{de}^{ca}+\epsilon
^{cde}T_{de}^{ab}\;,  \notag \\
\overline{W}_{abc} &=&W^{abc^{\ast }}=\epsilon _{ade}T_{bc}^{de}+\epsilon
_{bde}T_{ca}^{de}+\epsilon _{cde}T_{ab}^{de}\;,  \notag \\
X_{b}^{a} &=&(X_{a}^{b})^{\star }=iT_{cb}^{ac}\;,\;X_{a}^{a}=0\;.
\label{project}
\end{eqnarray}
The tensor $W^{abc}$ is fully symmetric in $abc$, as is $\overline{W}_{abc}$%
. The $\epsilon $ symbols are fully antisymmetric with $\epsilon
^{123}=\epsilon _{123}=1$. Incidentally, the octet components $X_{r}$ of $%
X_{b}^{a}$ can be easily expressed in terms of $T_{rs}$:
\begin{eqnarray}
X_{r} &=&\frac{i}{2}(\lambda _{r})_{ba}T_{cb}^{ac}=\frac{i}{2}(\lambda
_{r})_{ba}(\lambda _{s})_{ac}(\lambda _{t})_{cb}T_{st}  \label{comp8} \\
&=&\frac{i}{2}Tr\{\lambda _{r}\lambda _{s}\lambda
_{t}\}T_{st}=i(d_{rst}+if_{rst})V_{t}=-f_{rst}V_{st}\;.  \notag
\end{eqnarray}

Now we reconstitute $T_{cd}^{ab}$ from $W$, $\overline{W}$ and $X$. As a
first step we easily obtain:
\begin{equation}
\begin{array}{l}
\epsilon _{cde}W^{abe}=3(T_{cd}^{ab}+T_{cd}^{ba})+\delta
_{c}^{a}T_{de}^{eb}+\delta _{c}^{b}T_{de}^{ea}-\delta
_{d}^{a}T_{ce}^{eb}-\delta _{d}^{b}T_{ce}^{ea}\;, \\
\epsilon ^{abe}\overline{W}_{cbe}=3(T_{cd}^{ab}-T_{cd}^{ba})-\delta
_{c}^{a}T_{de}^{eb}+\delta _{c}^{b}T_{de}^{ea}-\delta
_{d}^{a}T_{ce}^{eb}+\delta _{d}^{b}T_{ce}^{ea}\;.
\end{array}
\label{comp10}
\end{equation}
The first expression is symmetric in $ab$ and antisymmetric in $cd$, while
in the second it is the other way around.\newline
If we now add these two results and bring in the tensor $X_{b}^{a}$, we get
the final formula:
\begin{equation}
T_{cd}^{ab}=\frac{1}{6}\epsilon _{cde}W^{abe}+\frac{1}{6}\epsilon ^{abe}%
\overline{W}_{cde}+\frac{i}{2}(\delta _{d}^{a}X_{c}^{b}-\delta
_{c}^{b}X_{d}^{a})\;.  \label{total}
\end{equation}
From here, via eq. (\ref{tensor}), we can obtain $T_{rs}$.

\bigskip
\begin{large}
\textbf{\ Acknowledgements}
\end{large}

One of the authors (NM) expresses grateful thanks to G.Marmo and G.Morandi
for gracious hospitality at the Universities of Napoli and Bologna, where
this work was initiated.

\end{document}